# Shock responses of nanoporous aluminum by molecular dynamics simulations

Meizhen Xiang[1], Junzhi Cui[2], Yantao Yang[2], Yi Liao[1], Kun Wang[1], Yun Chen[3], Jun Chen[1,4*]

[1]*Laboratory of Computational Physics, Institute of Applied Physics and Computational Mathematics, Beijing, 100088, China*
[2]*LSEC, ICMSEC, Academy of Mathematics and Systems Sciences, CAS, Beijing, 100090, China*
[3]*College of Computer and Information Sciences, Fujian Agriculture and Forestry University, Fuzhou,350002,China*
[4]*Center for Applied Physics and Technology, Peking University, Beijing 100071, China*

**Abstract:** We present systematic investigations on the shock responses of nanoporous aluminum (np-Al) by nonequilibrium molecular dynamics simulations. The dislocation nucleation sites are found to concentrate in low latitude region near the equator of the spherical void surfaces. We propose a continuum wave reflection theory and a resolved shear stress model to explain the distribution of dislocation nucleation sites. The simulations reveals two mechanisms of void collapse: the plasticity mechanism and the internal jetting mechanism. The plasticity mechanism, which leads to transverse collapse of voids, prevails under relatively weaker shocks; while the internal jetting mechanism, which leads to longitudinal filling of the void vacuum, plays more significant role as the shock intensity increases. In addition, an abnormal thermodynamic phenomenon (i.e., arising of temperature with pressure dropping) in shocked np-Al is discovered. This phenomenon is incompatible with the conventional Rankine-Hugoniot theory, and is explained by the nonequilibrium processes involved in void collapse. The influences of void collapse on spall fracture of np-Al is studied. Under the same loading velocity, the spall strength of np-Al is found to be lower than that of single-crystal Al; but the spall resistance is higher in np-Al than in single-crystal Al. This is explained by the combined influences of thermal dissipation and stress attenuation during shock wave propagation in np-Al.

**Keywords: A** dislocations; **B** porous material; void collapse; aluminum; shock.

*Corresponding author: jun_chen@iapcm.ac.cn; xiang_meizhen@iapcm.ac.cn.

# 1. Introduction

Nanoporous metals are a rapidly growing class of porous materials. Research on nanoporous metals is driven by their potential applications in various fields, such as catalysis, sensors, actuators, energy storage, blast protection, *etc*. [1]. Nanoporous metals can be prepared by dealloying and optical exposure [2,3]. Recent researches found that mechanical properties of on nanoporous are dependent on the porosity, the morphology such as ligament shape and connectivity and length scale of the pores and the ligaments [4-5]. The microscale deformation mechanism and continuum constitutive modeling of porous materials are widely studied in the literature [6-14]. These issues are fundamentally connected to ductile damage in metals through nucleation, growth and coalescence of voids .

The shock responses of porous metals have attracted great attention motivated by the need to explore methods to attenuate the effects of impact and blast. The testing and modeling of porous materials under shock is challenging. The pressure-density or shock velocity-particle velocity Hugoniot data for a large number of metals and porous substances have been obtained from experimental shock velocity data by impedance mismatch processing method [15]. The Hugoniot test results were used as a basis for derivation of numerous equation of states (EOSs) based on the jump approximation at the shock front and equilibrium state approximation behind the shock front [16]. It was of great interest to researchers that at moderate pressures of the order of several gigapascals, the derived Hugoniots for highly porous materials showed anomalous behavior resulting in pressure increasing with density dropping [16]. In order to account for the abnormality of test results, high temperature excitation mechanisms, such as the transition of the solid constituent of a powder into an electronic gas, are considered [17]. The shock waves in porous materials usually show multistep structures. Thus, a complete temporal record of stress or velocity is required for obtain proper information about the state being achieved during shock loading. However, such experimental information about the kinetic behavior is not readily available and is known for quite a limited range of porous materials, porosities, and loads. For material modeling, nonequilibrium effects would play significant roles for describing constitutive behaviors of porous materials under shock loading [16,18]. Unfortunately, the nonequllibrium processes triggered by the interactions between shock waves and the voids can not be directly observed by current experiment techniques.

Molecular dynamics (MD) simulation is an important alternative approach for studying mechanical behaviors of nanoporous materials, see the review in Ref.[19]. MD simulations can provide detailed mechanisms which cannot be directly observed in experiments. The micromechanisms obtained from MD simulations can be incorporated into constitutive equations for simulations at larger scales based on continuum mechanics. Nonequilibrium molecular dynamics simulations (NEMD) have been conducted to study the shock response of nanoporous metals in the past decade. Early NEMD simulations focus on collapse of single nanovoid under shock loading[20-23]. These studies have revealed the emission of shear loops from void surface and formation of hot spots. Recent NEMD simulations have studied the shock behaviors of np-copper with periodical or random distribution of nanovoids [24-29]. It is found that the void structures (shape, close/open, filled/empty) have great influences on elastic-plastic deformation, Hugoniot states, void collapse, hot spot generation, nano-jetting, melting and vaporization of np-copper.

This work presents systematic investigations on the shock responses of nanoporous aluminum (np-Al) by NEMD. We focus on several issues that are not studied in-depth in previous MD simulations, including void collapse mechanisms, thermodynamics characteristics and spall fracture of np-Al. For the void collapse mechanisms, we address the distribution of dislocation nucleation sites on the void surfaces. A continuum wave reflection theory and a resolved shear stress model are proposed to explain the distribution of dislocation nucleation sites. We also concern the competition between the plasticity mechanism and internal jetting mechanism of void collapse, which results in complex shape evolution during collapse. Thermodynamic features of wave propagation in np-Al are studied and an abnormal phenomenon, i.e., arising of temperature with pressure dropping is captured and explained. The differences between np-Al and single-crystal aluminum (sc-Al) in plastic strain relaxation and the corresponding dislocation activities (multiplication, annihilation and reaction) are discussed. Note that, previous MD simulations on shock behaviors of np-metals mainly focus on shock compression behaviors [13-22]. In the present work, the dynamical behaviors during both the compression and the release stages are addressed. The spall fracture of np-Al is also studied, with emphasis on spall strength and spall resistance of np-Al, with comparing to that of sc-Al.

The article is organized as follows. The sample model, simulation methods and post-processing algorithms are described in section 2. The detailed simulation results and discussions are presented in section 3. And the work is summarized in section 4.

## 2. Methodologies and notations

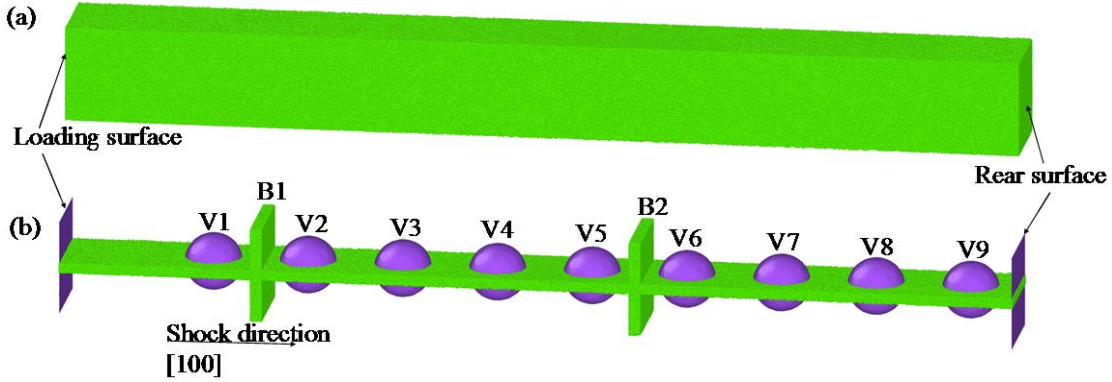

**Fig. 1.** The simulation model. (a) The out-side view of the simulated samples; (b) Illustration of the voids V1, …,V9; B1, B2 are two void-free material slices locating between the voids.

The atomic interactions in Al are described by an accurate embedded atom method (EAM) potential developed by Winey *et al.*[30]. The validity of the EAM potential under strong shock conditions is confirmed by comparing Hugoniot curves (*P-V* and *P-T* curves) and the melting curve with experimental data in our previous work [31]. The MD simulations are performed by the open source software LAMMPS [32]. The void-free single-crystal Al target is a sample with $500 \times 50 \times 50$ elementary cells and consists of 5032590 atoms. The [100], [010], and [001] crystallographic orientations are in accordance with the X1, X2, and X3 coordinate axes of the 3D Cartesian coordinate system. The sc-Al sample is relaxed by energy minimization using conjugate gradient method, followed by equilibration using NPT ensemble at 300K temperature and 0 GPa pressure for 50 ps. The np-Al samples are generated by deleting atoms in 9 equal-sized spherical voids from the sc-Al sample. The outside view of the samples are displayed in Fig.1 (a). The distribution of the 9 voids in the interior of the samples is illustrated by V1, …, V9 in Fig.1(b). Two void-free material slices (labeled as "B1" and "B2" in Fig.1(b)) are chosen for thermodynamic history analysis (as presented in section 3.2). The chosen material slices are vertical to the loading direction and are 3.0 nm thick. B1 locates between voids V1 and V2; while B2 locates between voids V5 and V6. Three np-Al samples are generated with the void radius being *R=5a,* 10*a* and 15*a,* respectively*,* where *a=*4.0248 angstrom is the lattice constant of the FCC aluminum. The distance between neighboring voids is set to be *L=*50*a,* the same as the side length of the lateral section of the sample. Thus, the porosities of the three np-Al samples ($f = {4\pi R^3}/{3L^3}$) are 0.4%, 3.4%, and 11.3% respectively. Note that, the left part of the sample with a length of 100*a* is bulk solid Al material free of voids. After the voids are generated, energy minimization and NPT equilibration at 300K and 0 GPa are carried out again for 50 ps for each np-Al sample.

The shock responses of the np-Al samples are studied by NEMD. Simulations on the sc-Al sample are conducted for comparison purpose. Shock wave is generated by moving a rigid Al piston with velocity $u_p$ along the [100] crystallographic orientation to the target sample for 20 ps. After that, the piston is removed to model the unloading process. Along the shock loading direction, free boundary conditions are applied. Periodic boundary conditions are applied to the X2 and X3 directions. Similar loading method is widely adopted in shock simulations [33,34]. The time-step in NEMD simulations is chosen to be 0.2 fs to ensure numerical stability. The velocity-Verlet algorithm is adopted to solve the MD equations. For each sample, we have conducted 3 simulations with the piston velocities being $u_p$ = 0.5 ,1.0 and 2.0 km/s, respectively. Thus, totally 12 simulations are performed.

The adaptive common neighbor analysis (ACNA) is adopted to identify the local lattice structure [35]. The dislocations are recognized by the dislocation extraction algorithm (DXA) based on the fundamental concept of Burgers circuit [36]. The DXA has been implemented into the open visualization tool OVITO as a standard modifier [36]. As part of the dislocation identification process, the DXA also constructs the defect surfaces, including void surfaces. Thus, it is very convenient to figure out the shape evolution of nanovoids during collapsing by using the DXA modifier in OVITO. For given material slice, the DXA analysis is conducted and the total length of the dislocation lines obtained as $L_d$. The dislocation density is calculated as $\rho_d = L_d / V$, with *V* being the volume of the material slice.

For the local atomic strain calculation, we firstly calculate the local deformation gradient *F* by a high-order finite element interpolation method, which is recently proposed by Yang *et al.*[37]. In this method, a FCC unit cell is viewed as a basic deformation element, with all the atoms in the unit cell are viewed as the nodes of the deformation element. Thus, the structure features of the FCC lattice is accurately represented by the basic deformation element. Nonlinear finite element interpolation is used to approximate the displacement field and strain field. Here, we briefly review the computational expressions of this method. A FCC unit cell contains 14 atoms (nodes), including 8 vertexes $P_1 \sim P_8$, and the 6 face center points $P_9 \sim P_{14}$. Denote the FCC unit cell in the initial configuration as $\Omega$. By the following affine transformation, $\Omega$ is transformed to a standard cubic cell $\Lambda$ with side length 2,

$$\xi(X) = \frac{X - X_c}{a}, \quad \forall X \in \Omega, \tag{1}$$

where $\xi$ denotes the coordinates in the standard cubic $\Lambda = [-1,1]^3$, $X_c$ denotes the center position of the FCC unit cell. The local deformation map from the initial configuration to the current (deformed) configuration is described by finite element interpolation,

$$x = \Psi(X) = \sum_{I=1}^{14} x_I \psi_I(\xi(X)) \tag{2}$$

where $x_I$ denotes the position of node $I$ in current configuration and $\psi_I$ are the finite element shape functions of the standard deformation cube, which take the following form,

$$\psi_I(\xi) = \begin{cases} \frac{1}{8}\prod_{j=1}^{3}(1+q_{Ij}\xi_j) - \frac{1}{8}\sum_{k=1}^{3}\prod_{j=1}^{3}(1-\xi_j^2)/(1-q_{Ik}\xi_k) & I = 1,\ldots,8, \\ \frac{1}{2}\prod_{j=1}^{3}(1-\xi_j^2)/\prod_{j=1}^{3}(1+q_{Ij}\xi_j) & I = 9,\ldots,14, \end{cases} \tag{3}$$

Where $q_{Ij}$ denotes the $j$-th component of the coordinates of node $I$ in the standard cubic. Then the deformation gradient in the FCC cell at position X is calculated by:

$$F(X) = \frac{\partial x}{\partial X} = \sum_{I=1}^{14} \frac{x_I}{a} \otimes \nabla \psi_I(\frac{X - X_c}{a}). \tag{4}$$

In large deformation problems, the Green-Lagrangian strain tenor is then calculated by $E = \frac{1}{2}(F^T F - I)$. The local hydrostatic strain invariant is obtained, $e_h = \text{Tr } E$. And the von Mises shear strain invariant is calculated as $e_s = \sqrt{\frac{1}{2}\text{Tr}[(E - e_h I)(E - e_h I)]}$.

where Tr[·] is the trace operator of matrix. The strain calculating technique is implemented into MD post-process package developed by the authors.

To obtain time histories of thermodynamic quantities of material elements is critical for understanding shock responses of materials. Usually, in shock experiments, the dynamical stress histories can be monitored by piezoelectric pressure sensors or manganin sensors embedded in the sample [38]. Such an embedded gauge is actually a Lagrangian station tracing the stress state during loading. One detailed publication on this topic is an article by Seaman[39], who describes how to recover the material response, using such Lagrangian data. To track the thermodynamic histories of stress and temperature of given material elements in MD simulations, we adopt an updated averaging approach. A material element $B$ in the initial configuration is represented by,

$$B = \{X \in R^3 : |X_1 - X_{c1}| \leq L_1, |X_2 - Y_{c2}| \leq L_2, |X_3 - X_{c3}| \leq L_3\} \quad (5)$$

where $X_c = (X_{c1}, X_{c2}, X_{c3})$ represents the center of the bin and $(L_1, L_2, L_3)$ represents the width of the bin in the three axes directions. The atoms in the sample is denoted as $\Gamma$, and the atom set in the material bin is represented by $S(B) = \{I \in \Gamma | X_I \in B\}$. To obtain the stress and temperature histories for the material element, we firstly calculate position of the center of the atom set in the current configuration by $x_c(B) = \frac{1}{N(B)} \sum_{I \in S(B)} x_I$, where $N(B)$ is the number of atoms in the initial bin $B$, $x_I$ is the position of atom $I$ in the current configuration. Average deformation gradient of the material bin $B$ is calculated by

$$\bar{F}(B) = \int_B F(X) \mathrm{d}X$$

where $F(X)$ is the deformation gradient field calculated by Eq.(4). Then, in the current configuration, the spatial domain obtained by the material bin B is determined from a continuum perspective as follows,

$$B^* = \{x_c(B) + \bar{F}(B) \cdot (X - X_c(B)) : X \in B\}. \quad (6)$$

After the spatial domain $B^*$ is determined, the Cauchy stress tensor and temperature of material bin by averaging the virial of atoms in $B^*$ in the current configuration,

$$\sigma(B) = \frac{1}{V(B^*)} \left( \sum_{I \in S(B^*)} \sum_{J \in \Gamma} \frac{1}{2}(x_J - x_I) \otimes f_{IJ} - m \sum_{I \in S(B^*)} \tilde{u}_I \otimes \tilde{u}_J \right)$$

$$T(B) = \frac{m}{3\kappa N(B^*)} \sum_{I \in S(B^*)} \tilde{u}_I \cdot \tilde{u}_I \quad (7)$$

where $S(B^*) = \{I \in \Gamma | x_I \in B^*\}$ is the atom set in the domain $B^*$, $N(B^*)$ is the number of atoms in $S(B^*)$. $V(B^*)$ is the volume of the bin in the current configuration and is calculated as,

$$V(B^*) = \det \bar{F}(B) \, V(B) = \det \bar{F}(B) L_1 L_2 L_3. \quad (8)$$

According to Eq.(6) and Eq.(8), the position, shape and size of $B^*$ in the current configuration is determined according to the average atomic deformation gradient and flow of the materials in the material bin $B$ (in the initial configuration). From a continuum perspective, the materials in $B^*$ can be viewed as materials that initially locates in $B$. However, from an atomistic perspective, due to diffusion of the atoms, the atoms in $B^*$ in the current configuration may be not exactly the same as that initially in $B$, i.e., $S(B^*) \neq S(B)$. The atom set $S(B^*)$ which is chosen to calculate thermodynamic quantities of the material bin is updated at each time step to coincide with the deformed bin in the current configuration. Thus our binning analysis is an "updated Lagrangian binning analysis method". This method is different from the traditional Eulerian binning analysis method, which can not tract the deformation histories of materials in the sample; the method is also different from the total Lagrangian binning analysis method, which can not insure the continuity of spatial distribution of the atoms that are chosen for calculating thermodynamics quantities.

In the following discussions, the press tensor ($p = -\sigma$) is sometimes used for convenience. We not only concern the hydrodynamic pressure $\bar{p} = \frac{1}{3}\text{Tr}[p]$, but also concern the evolution of shear stress. As the loading method in our simulations mimic 1D strain problem, the off-diagonal components of the stress tensor are negligible. Thus the shear stress was determined by the following equation [40],

$$\tau = \frac{1}{2}\left(p_{11} - \frac{(p_{22} + p_{33})}{2}\right) \tag{9}$$

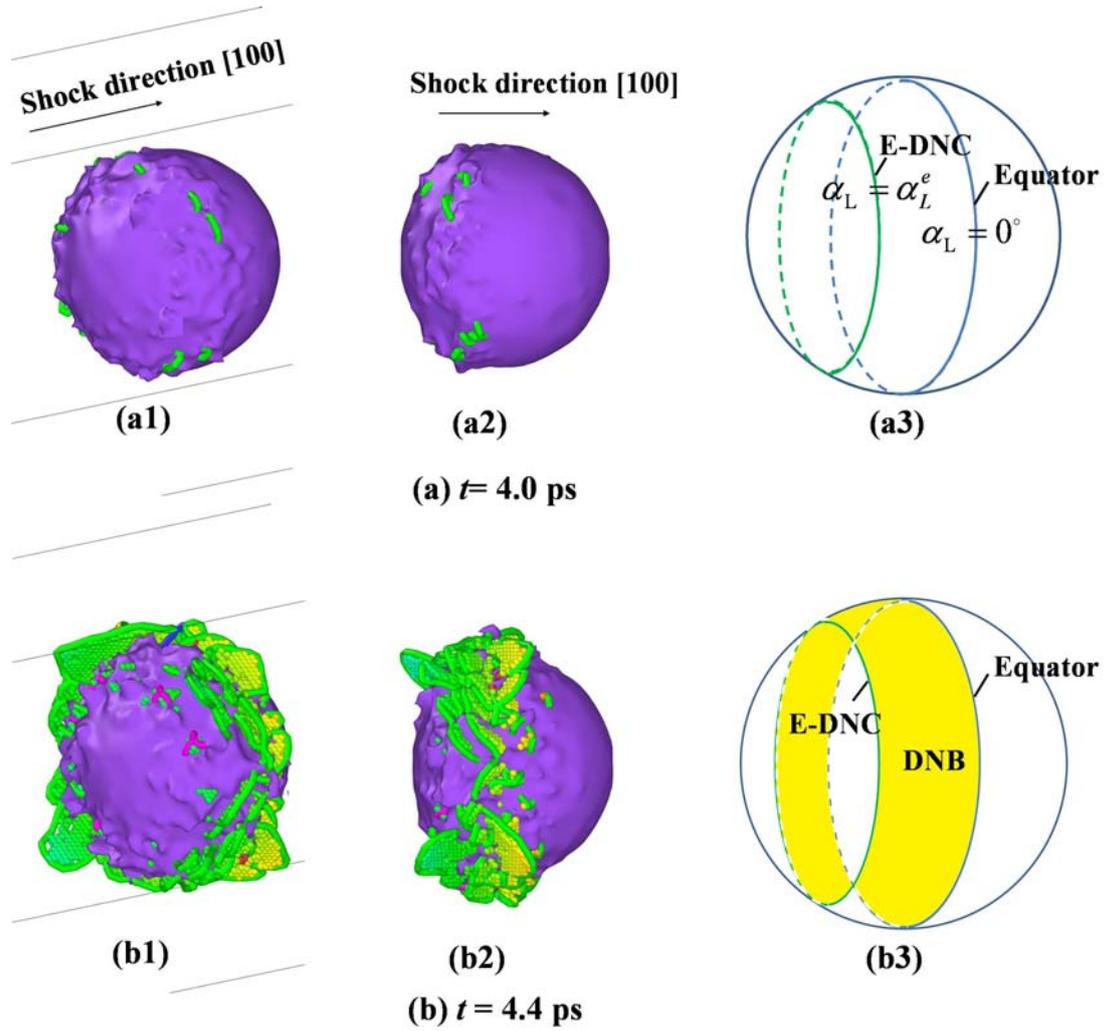

**Fig. 2.** Nucleation of dislocations from the surface of void V1 in the np-Al sample with $R=15a$ under $u_p = 0.5$ km/s, at times: (a) $t=4.0$ ps; (b) $t=4.4$ ps. The left figures (a1) and (b1) are orthogonal views; the middle figures (a2) and (b2) are front-side views; and the right figures (a3) and (b3) are sketched diagrams which illustrate the earliest appearing dislocation nucleation circle (E-DNC) and the dislocation nucleation band (DNB) on the void surface. The green lines in the left and middle figures are Shockley partial dislocations. The HCP atoms in the shear loops are displayed in (b1) and (b2).

## 3. Results and discussion

*3.1 Void collapse mechanisms: plasticity and internal jetting*

The dislocation nucleation process from the surface of void V1 in the np-Al sample with $R=15a$ under $u_p = 0.5$ km/s is displayed in Fig. 2. Upon the arrival of the shock wave at the void surface at $t=4.0$ ps, formation of the earliest dislocations are observed around a circle on the void surface at the shock side of the void, as shown in Fig. 2(a). For convenience, the great circle which is normal to the shock direction is referred to as the "equator"; And any circle on the void surface which is parallel to the equator is referred to as a "latitude circle". The latitude angle of a latitude circle is defined as the angle between the radius vector and the equator plane, and is denoted as $\alpha_L$, as illustrated in figures Fig. 2(a3) and Fig. 2(b3). The earliest appearing dislocation nucleation circle (E-DNC) is normal to shock direction and thus is a latitude circle of the nanovoid V1. The latitude of the E-DNC is about $\alpha_L = \alpha_L^e \approx 43.7°$ in this case. As the shock wave passes the void, shear loops nucleate in a band of low latitude angle ($0° \leq \alpha_L \leq \alpha_L^e$) on the void surface, between the E-DNC and the equator. This band is referred to as the dislocation nucleation band (DNB), as illustrated in Fig. 2(b3). In the high latitude region ($\alpha_L^e \leq \alpha_L \leq 90°$) on the void surface, nucleation of dislocations seldom occur. Note that, the dislocations are not symmetrically distributed at the two sides of the equator. In the early stage of void collapse, shear loops mainly nucleate on the front-side (the side which is shocked first) half void surface of the voids. Dislocation nucleation on the back-side half surface ($-90° \leq \alpha_L < 0°$) seldom occur until the materials impact the back-side half surface when the void is fully collapsed.

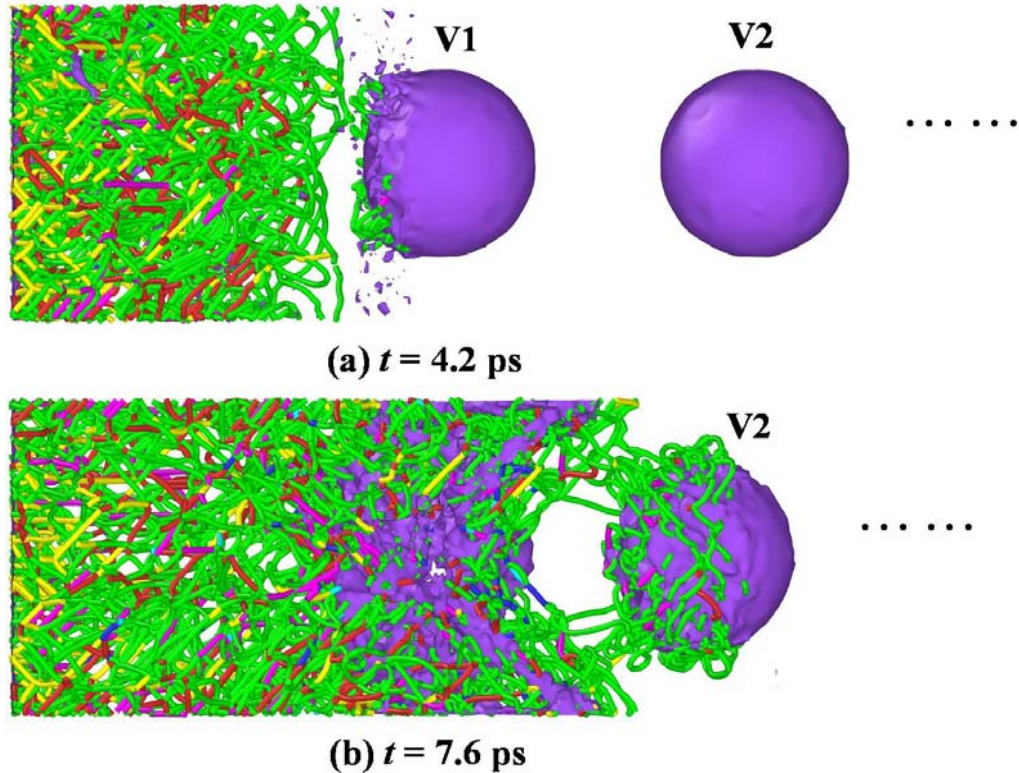

**Fig. 3.** Dislocations in the np-Al sample with $R=15a$ under $u_p = 1.0$ km/s, at times: (a) $t=4.2$ ps; (b) $t=7.6$ ps.

As the piston velocity increases to be $u_p = 1.0$ km/s, shock intensity is above the threshold for homogeneous dislocation nucleation in void-free bulk Al. In the np-Al samples, homogeneous dislocation nucleation occurs in the void-free bulk zone before the shock wave reaches the first void, as shown in Fig. 3(a). On the other hand, when the elastic precursor reaches the first void V1 at $t$=4.2 ps, dislocations immediately nucleate from the surface of void V1, as shown in Fig. 3(a). The latitude of the E-DNC is about $\alpha_L^e \approx 43.2°$ in the case of $u_p$ = 1.0 km/s, very close to that in the case of $u_p$ = 0.5 km/s. At $t$=7.6 ps, dislocations originated from void V2 meets with the dislocations originated from void V1, as shown in Fig. 3(b).

The MD simulations in present work confirmed two common features about the distribution of dislocation nucleation sites on the void surfaces in np-Al. Firstly, the latitude of the E-DNC $\alpha_L^e$ varies subtly, taking values from $40°$ to $45°$; Secondly, dislocation nucleation mainly occurs in low latitude region of the void surface between the E-DNC and the equator, with $0° \leq \alpha_L \leq \alpha_L^e$. This two features of the distribution of dislocation nucleation sites can be explained by continuum theories.

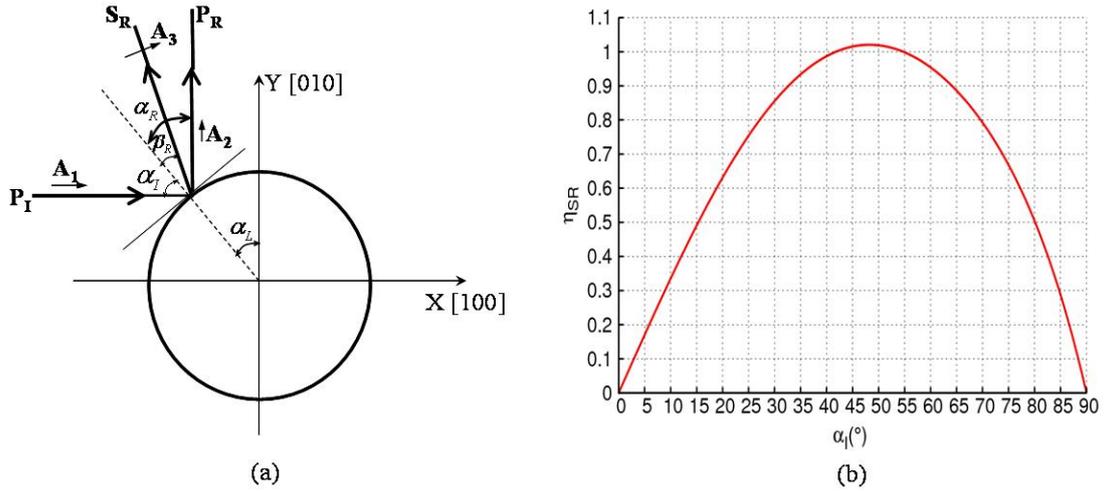

**Fig. 4.** Illustration of planar wave reflection on the surface of a spherical void. $P_I$ is the incident dilatational wave; $P_R$ is the reflected dilatational wave; $S_R$ is the reflected shear wave. $\alpha_I$ is the incident angle; $\alpha_R$ is the reflection angle of the dilatational wave; $\beta_R$ is the reflection angle of the shear wave. $A_1$, $A_2$ and $A_3$ are the amplitudes of displacements of $P_I$, $P_R$ and $S_R$.

From Fig. 2 and Fig. 3, the dislocation nucleation from void surfaces is triggered by the precursor wave. The amplitude of the precursor is below the dynamical yield point in bulk materials. Thus, before reaching the free surface of the first spherical void, the precursor wave can be roughly viewed as an elastic planar dilatational wave (also known as irrotational wave) traveling in bulk materials. The incident direction of the planar dilatational wave is oblique to the void surface at all points on the void surface except at the pole point ($\alpha_L = 90°$). From continuum wave theory, the oblique incident of planar wave would result in reflection of both dilatational wave and shear wave (also known as equivoluminal wave) [41]. Dislocation nucleation from the void surface is directly related to the amplitude of the reflected shear wave. Thus, the reflection coefficient for the shear wave is critical for understanding the dislocation nucleation features. Fig.4(a) illustrates the reflection of a planar dilatational wave on the spherical void surface. $D_I$ denotes the incident (dilatational) wave, $D_R$ denotes the reflected dilatational wave and $S_R$ denotes the reflected shear wave. The incident angle of $D_I$ is denoted as $\alpha_I$, and $\alpha_I + \alpha_L = 90^0$ holds; the refection angles of $D_R$ and $S_R$ are denoted as $\alpha_R$ and $\beta_R$. The amplitudes of the displacement perturbations of the incident dilatational wave $D_I$, the reflected dilatational wave $D_R$ and the reflected shear wave $S_R$ are denoted as $A_1, A_2$ and $A_3$. Reflection at a point on the spherical surface can be approximated by reflection at the tangent plane of that point. Under the assumption of isotropic elasticity and according to the free boundary conditions at the reflection surface, the reflection angles are related to the incident angle through the Snell law [41],

$$\begin{cases} \alpha_R = \alpha_I \\ \beta_R = \arcsin \dfrac{\sin \alpha_I}{\kappa} \end{cases} \quad (10)$$

where $\kappa = \sqrt{\dfrac{2(1-\nu)}{1-2\nu}}$ and $\nu$ is the Poisson ratio. And the reflection coefficients of $D_R$ and $S_R$ are obtained as [41],

$$\begin{cases} \eta_{DR} = \dfrac{A_2}{A_1} = \dfrac{\sin 2\alpha_I \sin 2\beta_R - \kappa^2 \cos^2 2\beta_R}{\sin 2\alpha_I \sin 2\beta_R + \kappa^2 \cos^2 2\beta_R} \\ \eta_{SR} = \dfrac{A_3}{A_1} = \dfrac{2\kappa \sin 2\alpha_I \cos 2\beta_R}{\sin 2\alpha_I \sin 2\beta_R + \kappa^2 \cos^2 2\beta_R} \end{cases} \quad (11)$$

From Eq. (10) and Eq. (11), the reflection coefficients are determined by the incident angle $\alpha_I$ and the Poisson ratio $\nu$. The nucleation of dislocations is directly induced by shear stress, thus is related to shear wave reflection coefficient $\eta_{SR}$. As for aluminum, the Poisson ratio is taken to be 0.342 and $\kappa$ is calculated to be 2.04 [42]. Fig.4(b) shows the relation between the shear reflection coefficient $\eta_{SR}$ and the incident angle $\alpha_I$. It is found that the shear reflection coefficient $\eta_{SR}$ reaches its maximum value 1.02 when $\alpha_I$ is $48.2°$, and the corresponding latitude is $\alpha_L \approx 41.8°$. As the precursor wave sweep the nanovoid in very short time, the precursor wave reaches different positions on the void surface almost at the same time. As a result, it is reasonable to assume that the E-DNC locates at the positions where reflected shear stress being the maximum. So, from the comtinumm prediction, the E-DNC would locate at latitude circle with $\alpha_L \approx 41.8°$. Our MD simulation results show that the latitudes of the E-DNC are about $40° \sim 45°$, which is very close to the prediction of the analytical model based on the continuum wave reflection theory.

After the precursor front passes the void, the stress field near the void surface may be roughly approximated by a biaxial static far-field compression problem. An approximated analytical solution of such a problem suggests that the tangential stress at the void surface is expressed as

$$\sigma_T = T_x[2(k-1)\cos(\pi - 2\alpha_L) + (1+k)], \qquad (12)$$

where $T_x$ is the far-field loading and $k$ is the biaxial loading ratio for uniaxial straincondition and is estimated to be $k = c_{12}/c_{11}$ [16]; And $c_{11}$=107.3 GPa, $c_{11}$=60.8 GPa are elastic constants for aluminum [43]. The relation between the resolved shear stress and the latitude is [23],

$$\tau_\theta = \frac{1}{2}T_x[2(k-1)\cos(\pi - 2\alpha_L) + (1+k)]\sin(2\theta - \pi + 2\alpha_L), \qquad (13)$$

where $\theta = 35.3°$ is the angle between the loading direction [100] and the {111} slipping planes. The tangential stress $|\sigma_T|$ reaches its maximum at the equator ($\alpha_L = 0°$), and the resolved shear stress $|\tau_\theta|$ reaches its maximum at latitude circle with $\alpha_L = 7.2°$, very close to the equator. That is why the dislocations nucleate favorably in low latitude region of the void surface instead of the high latitude region.

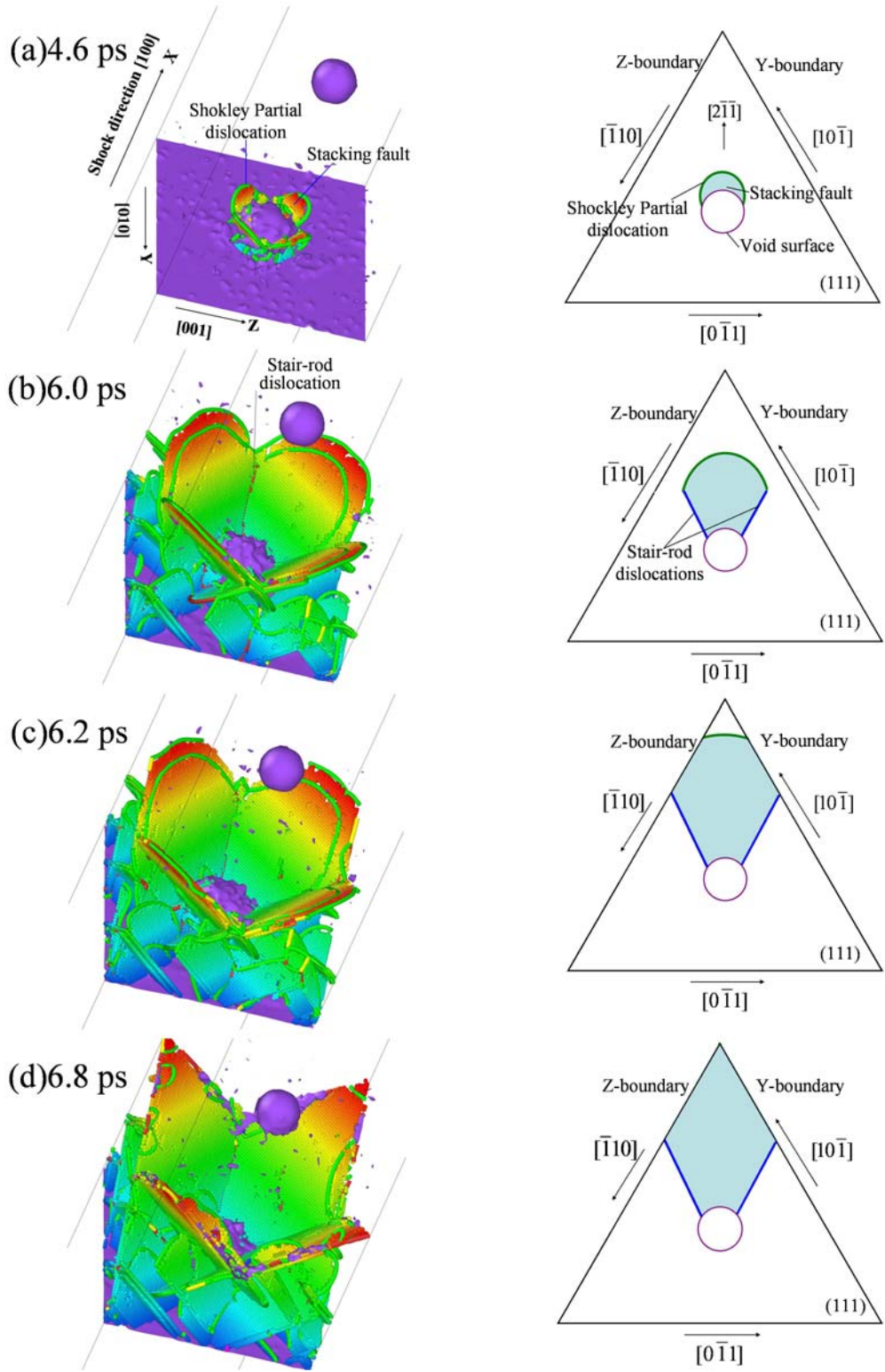

**Fig. 5.** Slipping of the shear loops nucleated from the surface of void V1 in the np-Al sample with $R=5a$ under $u_p = 1.0$ km/s, at times:(a) $t$=4.6 ps; (b) $t$=6.0 ps; (c) $t$=6.2 ps (d) $t$=6.8 ps. In each subfigure, the left figure displays the defects configuration from MD simulations; the right figure is the corresponding sketched diagram for the slipping of the shear loop on (111) plane with the leading Burgers vector being $\frac{1}{6}[2\bar{1}\bar{1}]$.

Fig.5 displays the expansion of four shear loops nucleated from the void surface of V1 in the np-Al sample with $R=5a$ under $u_p = 1.0$ km/s. In Fig.5, the atoms in the left-side figures are rendered by X positions to enhance the stereoscopic effects. The Y-boundary (Z-boundary) in the right-side figures is the intersection lines between the (111) slip plane and the periodic boundary of the simulation box in the Y (Z) direction. The Burgers vector of the leading partial dislocations of the four shear loops are $1/6[2\bar{1}\bar{1}]$, $1/6[2\bar{1}1]$, $1/6[211]$, $1/6[21\bar{1}]$; And the corresponding slip planes are $(111)$, $(11\bar{1})$, $(1\bar{1}\bar{1})$, $(1\bar{1}1)$. At the initial stage, the shear loops expand independently and the leading partial dislocations are in arc shapes, as shown in Fig. 5(a). As the shear loops expand farther, the leading Shockley partial dislocations on two different {111} slip planes meet and combine. This leads to formation of a stair-rod partial dislocation along one of the <110> directions, as shown in Fig. 5(b). The stair-rod dislocations are sessile and act as barriers to the glide of dislocations on the two {111} planes [44]. Note that only a part of a leading Shockley dislocation is involved in the reaction to form the stair rod; and the remained part of the leading Shockley partial dislocation keeps slipping, as shown in Fig. 5(b). At this stage, each shear loop is fan-shaped, with two stair-rods being the straight side edges and the remained leading partial dislocation being the arc edge of the shear loop fan. The four shear loops form four side-surfaces of a pentahedron with fourfold symmetry with respect to the loading direction. Similar defect configurations were also observed in other simulations [20-23]. The remained part of the Shockley partial dislocation continue to slipping until reaching the period boundaries of the simulation box in the Y and Z directions, as shown in Fig. 5(c). The Shockley partial dislocation is gradually absorbed by the boundaries. When the Shockley partial dislocation is completely absorbed by the sample boundaries, the shape of the shear loop becomes a rhombus, with the smaller angle between the adjacent sides being $60°$, as shown in Fig. 5(d).

In order to more intuitively show the void collapse process, Fig. 6 display the shape evolutions of the void V1 during collapse in the np-Al sample with $R=15a$ under $u_p =0.5$ km/s. As discussed above, the dislocations and shear loops mainly nucleate in low latitude region near the equator on the front-side half void surface in the case of $u_p=0.5$ km/s. The shear loops carry away the vacancies that comprise the vacuum of the void in the slipping direction. This leads to transverse "necking" of the voids, as illustrated in Fig. 6(a-d). The shape of the partially collapsed void is like a non-symmetry "dumbbell", with the front-side being smaller than the rear side. The non-symmetry feature is due to the non-symmetrical distribution of the dislocation nucleation sites at the two sides of the equator. Dislocations mainly nucleate at the front-side of the surface, and seldom nucleate at the back-side of the surface. So the front-side half surface collapses faster than the back-side half surface, leading to the non-symmetry shape of the partially collapsed void. At $t$=10 ps, the necked region is blocked by the collapsed materials and the original void is decomposed into two disconnected sub-voids, as shown in Fig. 6(e). After that, the two sub-voids shrink and finally disappear.

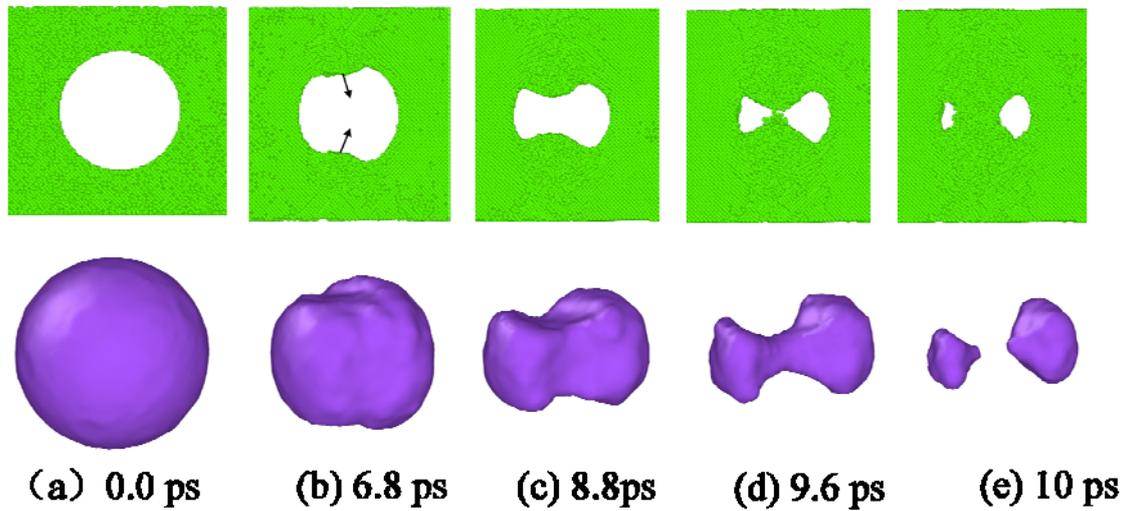

(a) 0.0 ps    (b) 6.8 ps    (c) 8.8ps    (d) 9.6 ps    (e) 10 ps

**Fig. 6.** The shape evolution of void V1 during collapse in the np-Al sample with $R$= 15$a$ under $u_p$=0.5 km/s. In each subfigure, the top figures are 2D side views of a selected section which is vertical to the Y direction and goes through the center of the sample; the bottom figures are the corresponding 3D perspective views of the surface of the partially collapsed void.

In addition to plastic mechanism, another mechanism of void collapse is internal jetting [25], to distinguish with the jetting from global surface of the sample. In our simulations, formation of internal jetting is obviously observed in stronger shock under $u_p$ =2.0 km/s, as shown in Fig.7. Internal jetting leads to filling of the void vacuum in the longitudinal direction, which is different from the plastic mechanism. The front-side half surface is firstly flattened by the shock wave, as shown in Fig. 7(b). Then the flattened front-side surface is gradually pushed into the back-side half of the void vacuum to form a groove, as in Fig. 7(c-e). The grooved void structure becomes thinner and thinner and becomes fully collapsed at 5.6 ps.

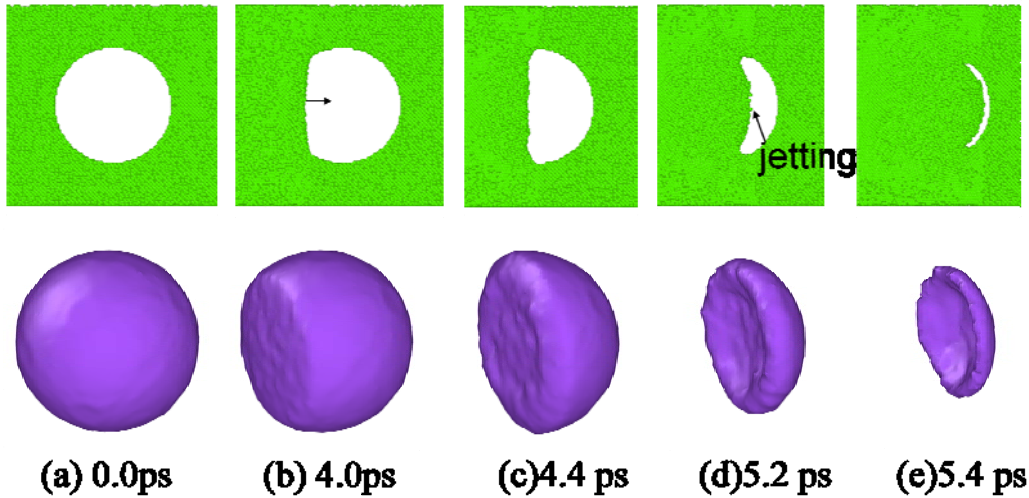

**Fig. 7.** The shape evolution of void V1 during collapse in the np-Al sample with $R=15a$ under $u_p = 2.0$ km/s.

It is well known that jetting is more obvious in shock melted metals, where shear stress is negligible. This indicates that internal jetting is mainly related to normal stress (pressure) instead of shear stress. On the other hand, plastic mechanism is related to the shear stress. This is the intrinsic difference between internal jetting mechanism and plastic mechanism. Generally, as the shock intensity increases, the resulted normal stress increases monotonically. However, the shear stress would not increases monotonically. In fact, in weaker shock regime, the shear stress may increases as the shock intensity increases. As the shock intensity exceeds some critical value, the shear stress would decrease as the shock intensity increases due to temperature softening of the yield strength. Especially, when the shock intensity approaches the shock melting point, shear stress would be negligible. This indicates that the plastic mechanism would prevail under weaker shock; and the internal jetting mechanism would play more significant roles as the shock intensity increases.

3.2 *Thermodynamic characteristics under compression and release*

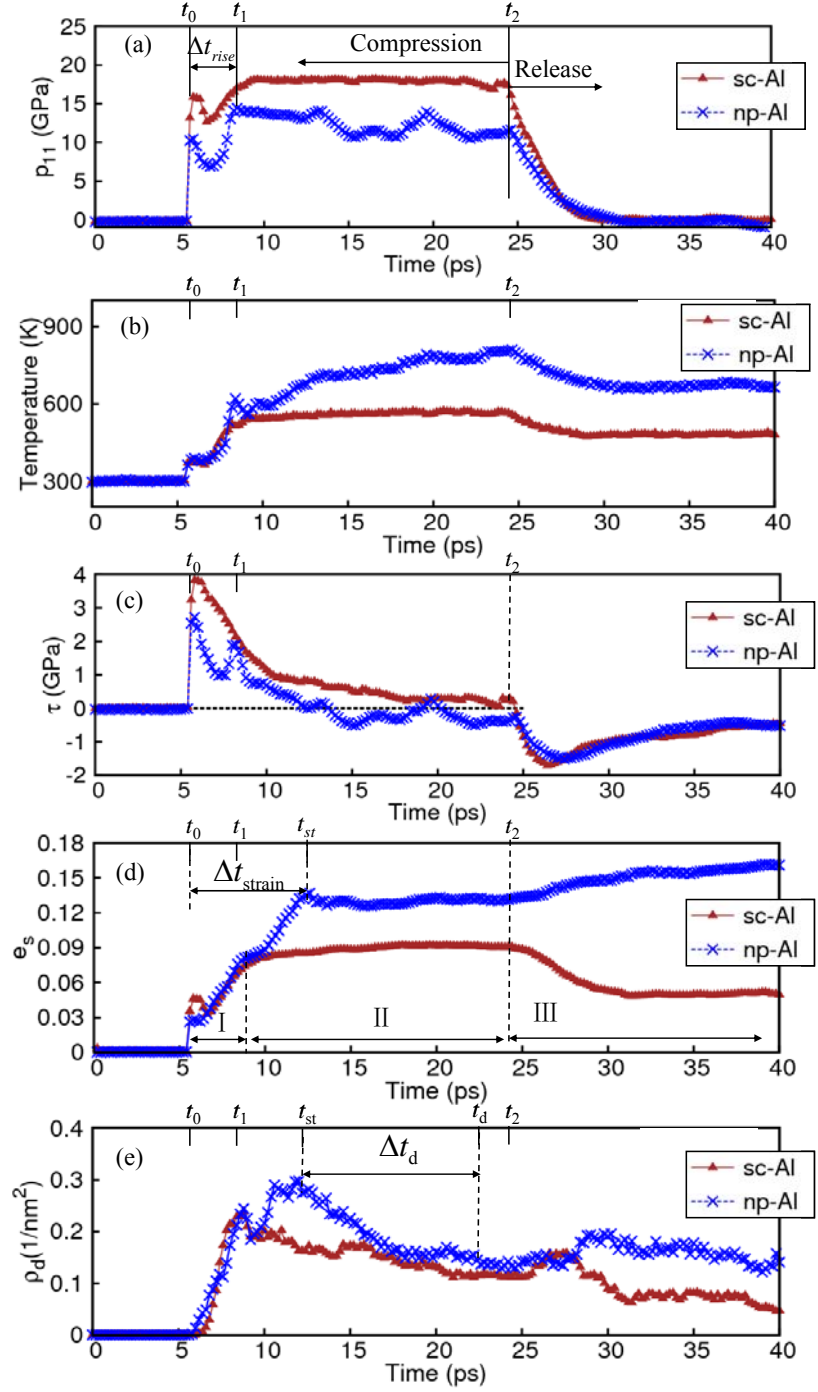

**Fig. 8.** Comparing of time history curves of material slice B1 in the sc-Al sample and the np-Al sample with $R=15a$ under $u_p=1.0$ km/s. (a) Press histories, and illustration of typical times $t_0$, $t_1$ and $t_2$ and the rise time $\Delta t_{rise}$ in the np-Al sample; (b) Temperature histories; (c) Shear stress histories; (d)The von Mises shear strain invariant histories and the shear strain relaxation time $\Delta t_{strain}$; (e) Dislocation density histories, and illustration of dislocation relax time $\Delta t_d$.

In Fig. 8(a-e), the temporal history curves of normal press ($p_{11}$), temperature, shear stress($\tau$), shear strain($e_s$) and dislocation density ($\rho_d$) of the material slice B1 in the np-Al sample with $R=15a$ are compared with that in sc-Al sample, under $u_p$=1.0 km/s. In the np-Al sample, the material slice B1 lies between void V1 and void V2, as illustrated in Fig. 1(b). The material slice B1 in the sc-Al sample locates at the same position as in the np-Al sample. According to the press history, three typical time instants, $t_0$, $t_1$, and $t_2$ are marked in Fig. 8(a) for the np-Al. $t_0$ is the beginning point of the wave front; $t_1$ is the ending point of the wave front; $t_2$ (~24.8 ps) marks the time instant when the release wave reaches the material slice B1. The release wave is originated from the loading surface by removing the piston at $t$=20 ps. From Fig.8(a), the amplitude of the main wave in the np-Al sample is 14.1 GPa, which is lower than that in the sc-Al sample (18.0 GPa).

From Fig. 8(a-b), there are significant differences in the evolution characteristics of press and temperature histories in sc-Al and np-Al. In sc-Al, the material reaches Hugoniot equilibrium in very short time after the passage of the wave front, and both the press and the temperature keep at constant values (the Hugoniot values) during the compression stage from $t_1$ to $t_2$ . However, in np-Al, both the press and temperature can not be equilibrated to the Hugoniot state on the time scale of tens of picoseconds. From Fig. 8(a), after the arrival of the main wave at $t_1$, the press starts to drop slowly with small fluctuations in the np-Al sample. On the other hand, the temperature in the np-Al sample keeps rising during the compression stage, as shown in Fig. 8(b). The arising of temperature with pressure dropping is an abnormal phenomenon if one assumes the applicability of the conventional Rankine-Hugoniot theory. From conventional Rankine-Hugoniot theory, the press and temperature would keep constant after the swept of the shock front. The abnormality of temperature arising with pressure dropping is due to the highly nonequilibrium processes involved in void collapse. Thus, more accurate constitutive models, which take the nonequilibrium effects into account, are required to describe the shock behaviors of nanoporous materials. The NEMD simulations can provide necessary details about the nonequilibrium processes.

To show more details about the nonequilibrium processes, evolution of local temperature field around voids V1 and V2 and the material slice B1 are displayed in Fig. 9. When the wave front reaches B1, the void V1 is partially collapsed, as shown in Fig. 9(a). Soon after the wave passes B1, it reaches the surface of void V2. Reflection on the surface of void V2 forms a rarefaction wave propagating toward B1 and leads to drop of the pressure. On the other hand, the collapsing of void V1 leads to strong impaction of materials on the back-side surface of void V1. This impaction forms a spherical compression wave propagates toward B1. The spherical compression wave leads to formation of a high temperature circular ring in B1, as shown in Fig. 9(b). When the void V1 is fully collapsed, a hot spot forms in the center of B1, as shown in Fig. 9(c). At this instant, the temperature field is strongly inhomogeneous. The collapse of void V2 also forms a hot spot. The average temperature of B1 is lower than the materials in the collapsed region of voids V1 and V2. Thus, thermal diffusion would lead to gradually rising of the average temperature of B1 (see Fig. 8(b)). As time goes on, the temperature field of B1 becomes more and more homogeneous due to thermal diffusion, as shown in Fig. 9(d).

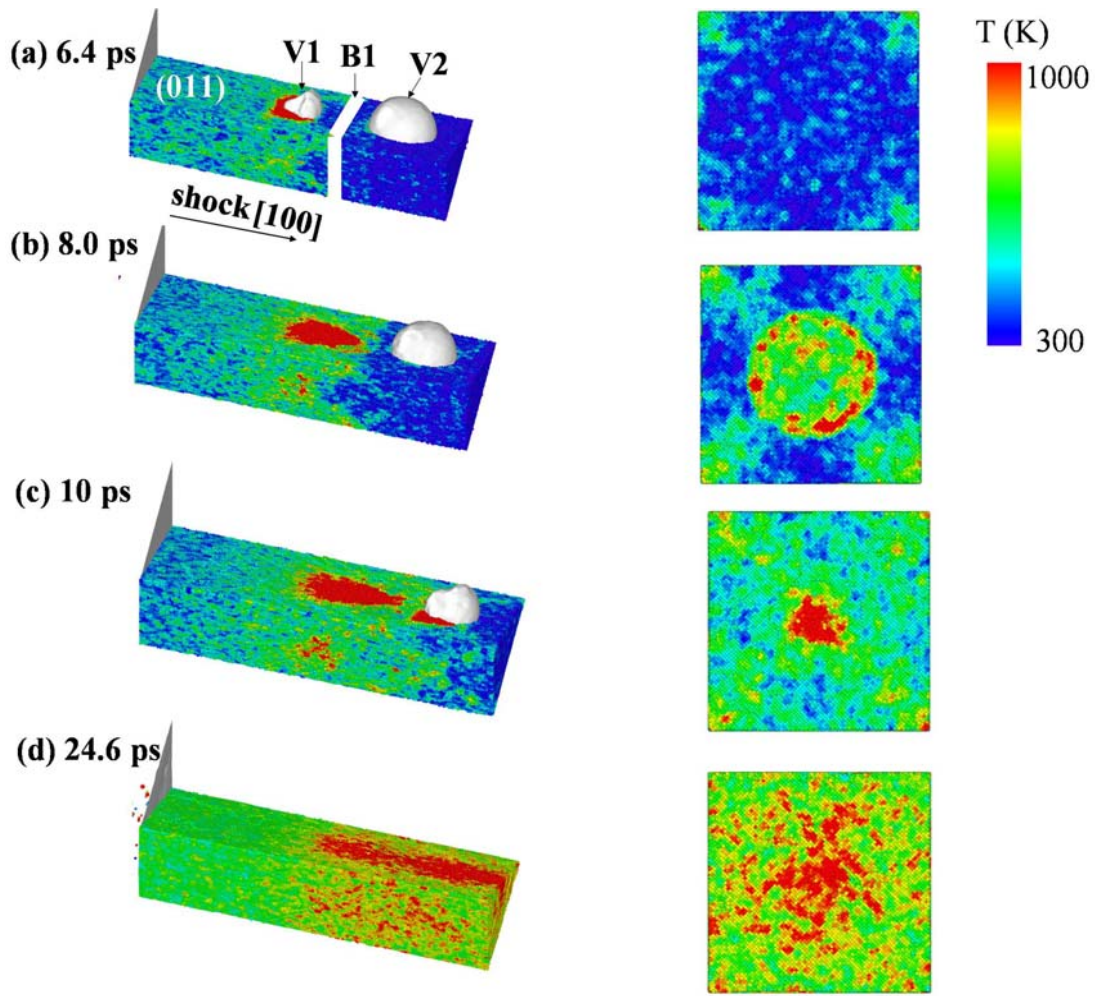

**Fig. 9**. The evolution of local temperature field around voids V1 and V2 and the material slice B1. The left figures are perspective views shown the temperature field around V1 and V2. The atoms are divided into two parts by a cut plane, only one part is shown. The cut plane is normal to [011] direction and goes through the center of the sample. The right figures are right-side views of the material slice B1.

Upon shock at $t_0$, the material slice B1 is firstly compressed in uniaxial(1D)

manner, and $p_{11} > p_{22} = p_{33}$, resulting in non-zero shear stress $\tau$ according to Eq.(9). At this instant, the shear stress sharply builds up to the peak value, similar to the press, as shown in Fig. 8(c). After that, transverse plastic flow leads to the release of shear stress behind the shock front (Fig. 8(c)). This process is referred to as the plastic relaxation process. During plastic relaxation, the strain evolves from 1D compression to 3D compression through dislocation slipping. And von Mises shear strain invariant $e_s$ (Fig. 8(d)) and the dislocation density $\rho_d$ (Fig. 8(e)) increase in this process. The 1D to 3D plastic relaxation behind the shock front was observed in X-ray diffraction experiments and was also confirmed in large scale MD simulations [45-46]. From Fig. 8(d), the shear strain approaches its saturated value soon after the passage of the wave front in both sc-Al and np-Al. At the same time, the dislocation density of B1 reaches maximum, as shown in Fig. 8(e). During the compression from $t_1$ to $t_2$, the shear strain keeps constant at the saturated value, while the dislocation density decreases due to effect of dislocation annihilation.

The differences in plastic relaxation between sc-Al and np-Al mainly lies in shear strain evolution during the release stage. After the release wave reaches B1 at $t_2$ (~24.8 ps), the shear stress $\tau$ of both the np-Al and sc-Al decreases to the negative regime, i.e., $\tau < 0$, as shown in Fig.8(c). This means that the release process is not a prompt 3D release process. The negative shear stress has to be relaxed gradually by shear slipping (plastic relaxation). In sc-Al sample, after the release wave arrives at B1 at $t_2$, the shear strain decreases, as shown in Fig.8(d). This is attributed to partial recovery of the existing crystal slipping bands which were produced during shock compression. Note that, although plastic strain accumulation at macroscales is irreversible, completely or partially recover from severe deformation is widely observed at nanoscales [23,47]. This is because that the spatial and temporal scales are too small for the dislocations to be well entangled at nanoscales. Thus, dislocation forest and junctions can not be largely produced as at macroscales. As a result, the shear slipping bands which are produced under shock compression are partially recovered during the release process. This leads to the drop of the average shear strain of B1 during release process in sc-Al.

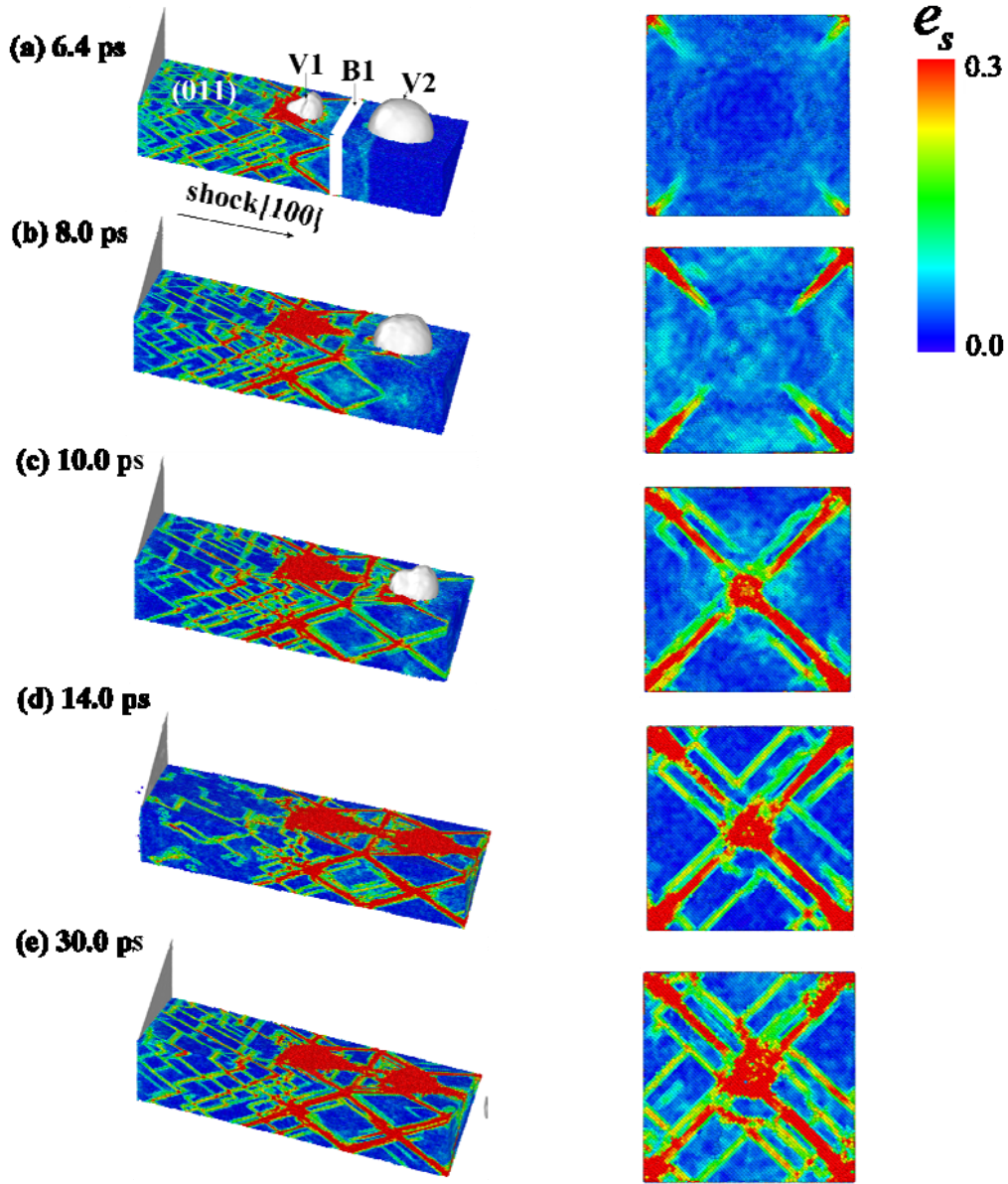

**Fig. 10.** The evolution of local shear strain field around voids V1 and V2 and the material slice B1. The left figures are perspective views. The right figures are right-side views of the material slice B1.

The dynamical evolution of shear strain in np-Al is quite different from that in sc-Al sample. To analyze the dynamical evolution of the shear strain in np-Al, we combine the shear strain history curve in Fig. 8(d) with the local shear strain field encoded atomistic configurations around voids V1 and V2 and the void-free materials B1, which are displayed in Fig. 10. The shear strain history in np-Al displays a three-stage accumulation process. The three stages are marked by Ⅰ, Ⅱ and Ⅲ in Fig. 8(d). In stage Ⅰ, shear strain history in np-Al coincides well with that in sc-Al. In this stage, the shear strain accumulation is attributed to plastic relaxation from 1D to 3D compression by motion of extended dislocations in the void-free material slice B1. In sc-Al, plastic relaxation is due to homogenous nucleation and motion of dislocations. In np-Al, the 1D to 3D relaxation of the void-free material slice B1 in stage I is attributed to the motion of dislocations which are nucleated from void surface V1 and then move into B1. The motion of dislocations form shear slipping bands, as shown in Fig. 10. The shear loops nucleated from the void V1 interact with the period boundaries before moving into the slice B1. As the result, the earliest shear slipping bands in the slice B1 appear at the four corners and grow into the interior along the diagonal lines, as shown in Fig. 10(a-b). After the full collapse of void V1, there is a short pause for the shear strain accumulation. This short pause is corresponding to the time interval required for the wave travels from void V1 to the void V2. After that short pause, the dislocations nucleated due to the collapse of void V2 move into the material slice B1, which leads to formation of new shear bands (see Fig. 10(c-d)) in the material slice and results in increase in the shear strain (see Fig. 8(d)). The shear slipping bands in the material slice and the average von Mises shear strain of the slice B1 saturates at $t_{st} \approx 12$ ps, as illustrated in Fig. 8(d). The time for the shear strain to increase from zero to the saturated value is referred to as the plastic strain relaxation time, and is denoted as $\Delta t_{strain}$, as shown in Fig.(d). After $t_{st}$, the shear strain keeps constant until the release wave arrives at $t_2$. As discussed above, the release process is a 1D to 3D process and 1D release induces negative shear stress $\tau$. The negatitive shear stress can be relaxed through the recover of the shear slipping in sc-Al sample. The recover of the shear slipping bands leads to drop of the accumulated shear strain. The recover of the shear slipping in sc-Al is due to lack of stable defect structures such as dislocation forests and junctions. In np-Al, however, the dislocations are severely entangled and much obstacles for dislocation motion are produced. This largely reduces the possibility of recover of the existing shear slipping bands. As a result, the negative shear stress can not be relaxed by recovery of existing slipping bands, and has to be relaxed by producing new slipping bands, as shown in Fig. 10(e). This leads to the increasing of von Mises shear strain during the shock release stage, as observed in Fig. 8(d).

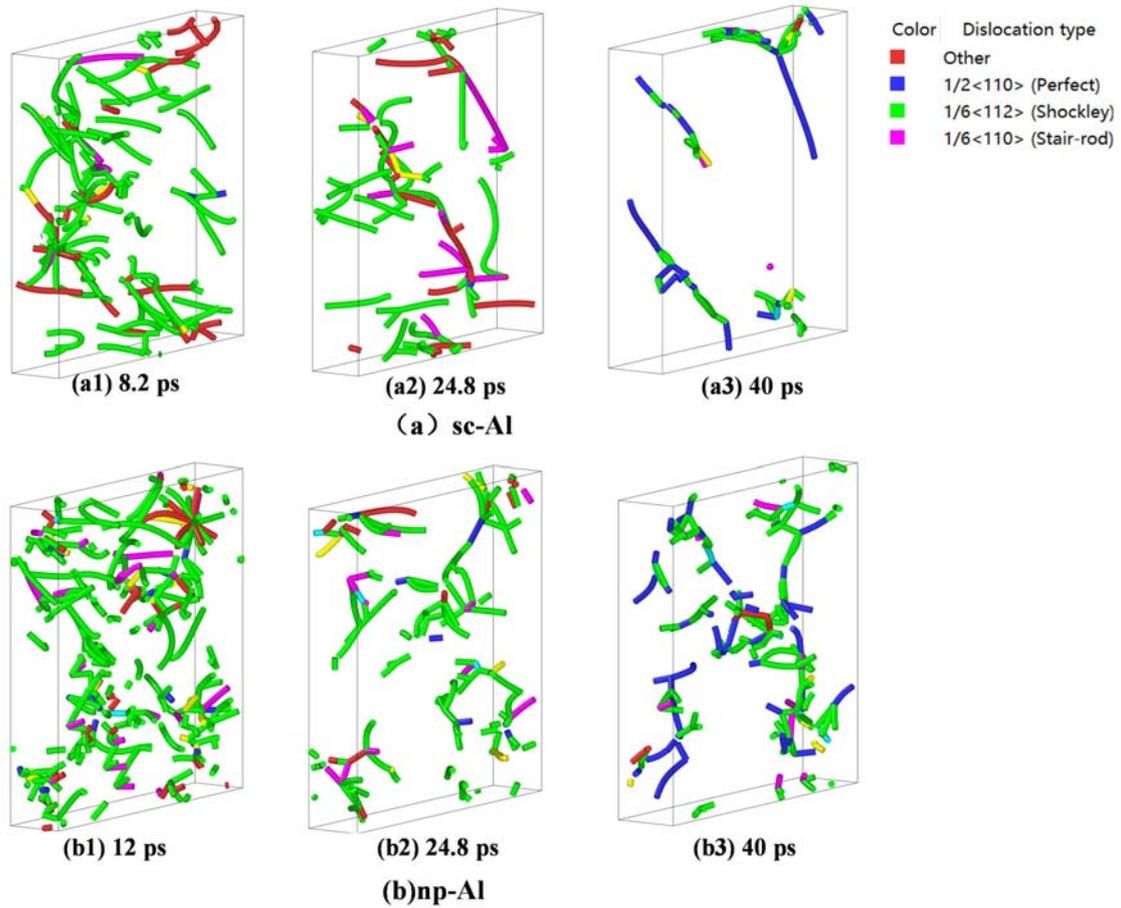

**Fig. 11.** The evolution of dislocation configurations in the material slice B1 at typical times in (a) sc-Al ample and (b) np-Al sample with $R=15a$, under $u_p=1.0$ km/s.

Fig. 11 shows the dislocation structures in the material slice B1 at several typical time instants in sc-Al and np-Al. The dislocation lines are colored according to their type. Most of dislocations nucleated homogeneously in the sc-Al sample or nucleated from the void surfaces in sc-Al sample are Shockley partial dislocations. Fig. 11(a1) and Fig. 11(b1) display the dislocation configurations when the dislocation densities of the slice B1 reach the maximum value, at $t=8.2$ ps for the sc-Al sample and $t=12.0$ ps for the np-Al sample (see Fig. 8(e)). At this instant, the amount of Shockley partial dislocations are much larger than all other types of dislocations in both sc-Al and np-Al. The dislocation configurations for the sc-Al and np-Al at $t=24.8$ ps, just before the arrival of the release wave, are displayed in Fig. 11(a2) and Fig. 11(b2) respectively. It is found that the dislocation lines become much more sparse during the compression stage, which is consistent with the history curves of dislocation density in Fig. 8(e). This explicitly confirm the annihilation of dislocations under compression state. It is found from Fig. 11(a1-a2) and Fig. 11(b1-b2) that, under compression, the Shockley partial dislocation largely disappears and amount of stair rod dislocations forms. The formation of the stair-rod dislocations is attributed to the intersection of shear loops on different slipping planes [44]. The dislocation configurations in the sc-Al and np-Al at $t$=40 ps, after the normal press has been released to zero, are displayed in Fig. 11(a3) and Fig. 11(b3) respectively. From Fig. 11(a2-a3), in the sc-Al sample, both the Shockley partial dislocations and stair-rod dislocations are largely annihilated during the release stage, while only very few perfect dislocations remain after the sweeping of the release wave. From Fig. 11(b2-b3), the total amount of the dislocation lines is not obviously changed during the release stage, which is consistent with dislocation density history curve in Fig. 8(e). However, the proportion of each dislocation type varies dramatically. The proportion of Shockley partial dislocations decreases from 81% to 63%; while the proportion of the perfect dislocations increases from 0.3% to 31%. This indicates that constriction of the extended dislocations to perfect dislocations occurs during the release stage.

According to elastic theory of dislocations, the width of an extended dislocation is approximated by $d = Gb^2 / 4\pi\gamma$, where $G$ is the shear modulus, $b$ is the magnitude of the Burgers vector and $\gamma$ is stacking fault energy. Thus, constriction of extended dislocations to perfect dislocation is energetically favorable in materials with high stacking fault energy. Aluminum is a typical FCC metal with high stacking fault energy (104-142 mJ m$^{-2}$) [48]. Due to the high stacking fault energy of Al, macroscale experiments on plastic behaviors of coarse-grained Al have confirmed the existence of perfect dislocations, with seldom evidence of Shockley partial dislocations and stacking faults [49,50]. On the other hand, both MD simulations and experiments have shown partial dislocations and stacking faults in nanocrystalline Al [48,51-52]. In our simulations, both partial dislocations and perfect dislocations are observed. Moreover, we have captured the whole dynamical process for the transformation of Shockley partial dislocations to stair-rods and perfect dislocations in both sc-Al and np-Al.

Stress attenuates as the shock wave propagates father into the np-Al sample. To confirm this, we have compared the stress histories of the material slices B1 and B2 in the np-Al sample with $R$=15$a$ under $u_p$=1.0 km/s. B1 locates between voids V1 and V2, while B2 locates between voids V5 and V6, as illustrated in Fig.1(b). The amplitude of the elastic precursor, the maximum press and the maximum shear stress of B1 are 7.2 GPa, 14.1 GPa and 2.8 GPa. These values have attenuated to be 4.1 GPa, 11.9 GPa and 1.4 GPa in the material slice B2.

3.3 *Global spall fracture of nanoporous aluminum*

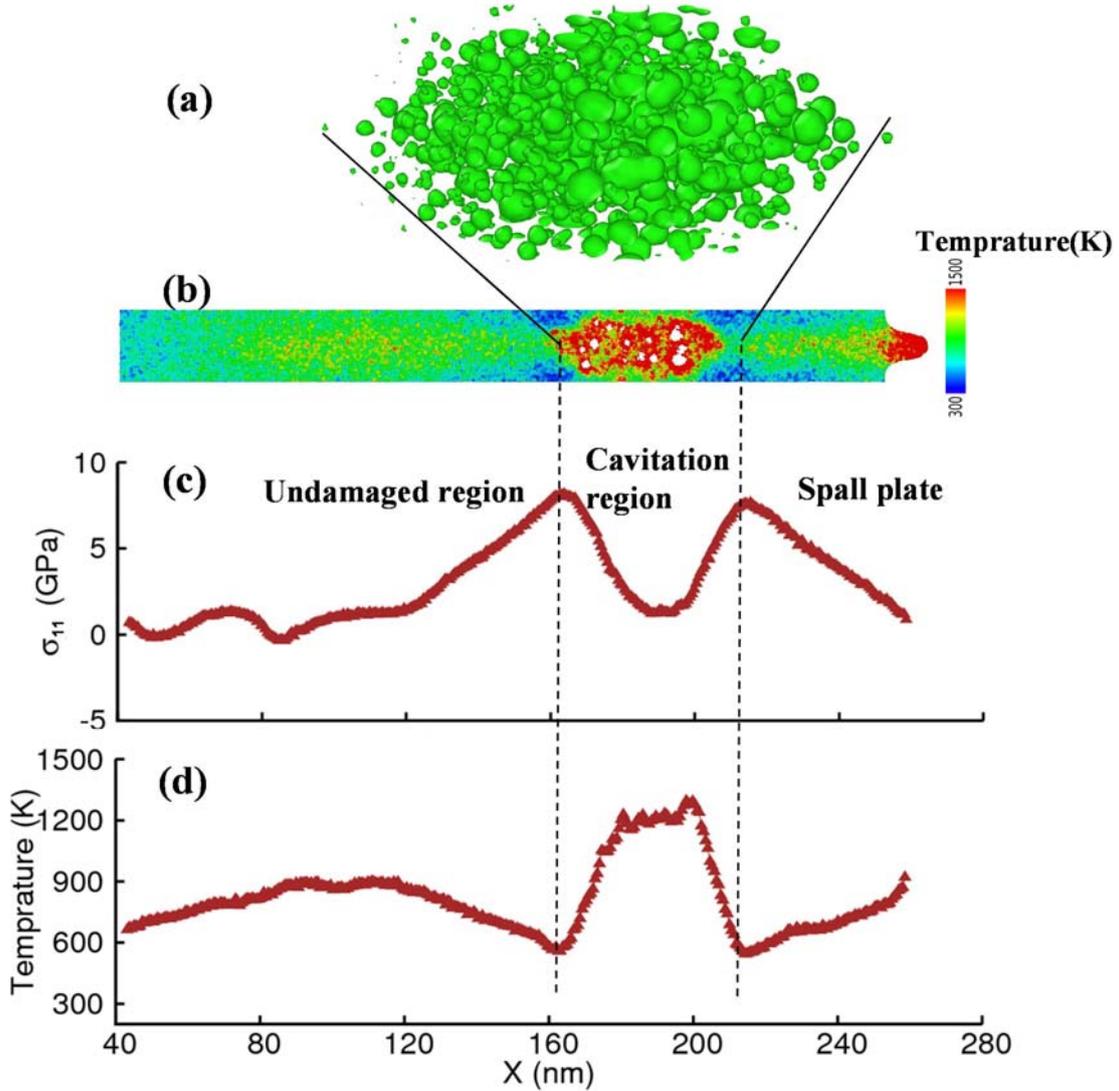

**Fig. 12.** The damage configuration and thermodynamic profiles of the np-Al sample with R=15a under up=2.0km/s, at t=45 ps. (a) The voids nucleated during spalling in th cavitation region; (b) Section view of the damaged sample with atoms colored by atomic temperature; (c) Normal stress profile;(d) Temperature profile.

When the loading piston is removed from the simulations at *t*=20 ps, a rarefaction wave is originated from the loading surface. In addition, when the incident compression wave reaches the rear free surface of the sample and reflects, another rarefaction forms. These two rarefaction waves propagate in opposite directions in the sample. When they meet, net tensile stress emerges. If the tensile stress in the loading direction $\sigma_{11}$ exceeds the spall strength $\sigma_{sp}$, spall fracture occurs through cavitation, i.e., nucleation, growth and coalescence of voids. For np-Al samples, when the incident shock reaches the free surface, the porous zones have undergone strong modifications due to void collapse. These processes would significantly influence the global fracture of the np-Al.

Fig. 12 displays the spall damaged configuration and corresponding stress and temperature profiles in the np-Al sample with $R=15a$ under $u_p=2.0$ km/s, at $t=45$ ps. The damaged sample can be divided into three regions according to spatial distribution of the nucleated voids, as illustrated by the two dotted lines in Fig. 12. These three regions are the undamaged main body region, cavitation region and spall plate region, as illustrated in Fig. 12. Void nucleation occurs at high tensile stress positions which bound the cavitation region, as shown in Fig. 12(a-b). In the interior of the cavitation region, tensile stress is released due to nucleation and growth of voids, as shown in Fig. 12(c). It is found that the average temperature of materials in the cavitation region is much higher than that of other two regions, see Fig. 12(d). This is attributed to thermal dissipation during cavitation.

Table 1. Spall parameters of the sc-Al sample and the np-Al samples under $u_p=1.0$ km/s and 2.0 km/s.

| $u_p$(km/s) | $R$ | $\sigma_{sp}$(GPa) | $T_{sp}$(K) | $\Delta T_{sp}$(K) |
|---|---|---|---|---|
| 1.0 | 0* | 9.06 | 283 | 454 |
|  | 5a | 9.01 | 295 | 494 |
|  | 10a | 8.63 | 303 | 588 |
|  | 15a | ---** | --- | --- |
| 2.0 | 0 | 9.87 | 318 | 221 |
|  | 5a | 9.83 | 326 | 381 |
|  | 10a | 9.35 | 365 | 579 |
|  | 15a | 8.46 | 551 | 647 |

Notes: * $R=0$ represents the sc-Al sample.
** In the np-Al sample with $R=15a$, spall does not occur.

The spall strength $\sigma_{sp}$ is calculated as the maximum tensile stress experienced by the sample during spalling [53]. The temperature just before void nucleation is defined as the spall temperature $T_{sp}$ [53]. The temperature arising during the void nucleation stage is defined as $\Delta T_{sp}$. These spall parameters in different simulations are listed in Table.1. It is found that, under the same loading velocity, the spall temperature $T_{sp}$ increases as the porosity increases. This is because that higher porosity would result in more heat production during void collapse under shock compression. Higher temperature leads to lower spall strength [53]. As a result, the spall strength of np-Al decreases as the porosity increases, as shown in Table.1.

To describe the shock protection capability of porous materials, the critical shock impact velocity that induces spall fracture is referred to as "spall resistance" for convenience. If impact velocity is below the spall resistance, spall (cavitation) does not occur; otherwise spall occurs.

A very interesting result obtained in our simulations is that spall does not occur in the np-Al sample with $R=15a$ under loading velocity $u_p=1.0$ km/s. However, spall occurs in all other np-Al samples and in the sc-Al sample under the same loading condition. This indicates that in all the samples simulated in the present work, the np-Al sample with $R=15a$ has the highest spall resistance, even it has the lowest density and the lowest spall strength. To understand this, two competing mechanisms of the influences of local void collapse on global spall fracture must be taken into account. The first mechanism is stress attenuation. As in porous materials, the local void collapse induces attenuation of stress. The stress attenuation is more remarkable for higher porosity sample. Thus, the maximum tensile stress experienced by the materials would be lower for higher porosity sample. On the other hand, higher porosity would result in lower spall strength due to temperature softening effect, as shown in Table.1. As porosity increases, whether the spall resistance increases or decreases is dependent on the competing between the stress attenuation mechanism and the spall strength softening mechanism. If the first mechanism (stress attenuation) prevails, the spall resistance increases; otherwise, if the second mechanism (softening) prevails, the spall resistance decreases. In the simulations of the present work, the stress attenuation mechanism prevails and the spall resistance of np-Al sample is higher than that of sc-Al sample.

**4. Summary**

This work has presented systematical investigations on shock responses of nanoporous Al by nonequilibrium molecular dynamic simulations. The void collapse mechanisms, wave propagation characteristics and spall fracture are concerned. The main conclusions are summarized as following:

(1) Two mechanisms of void collapse under shock compression are revealed, i.e., the plasticity mechanism and the internal jetting mechanism. The plasticity mechanism leads to transverse necking of voids and the internal jetting mechanism leads to longitudinal filling of the void vacuum. The competition between the plasticity mechanism and the internal jetting mechanism leads to complex shape evolution of the partially collapsed voids. The plasticity mechanism prevails under relatively weaker shocks, while the internal jetting mechanism plays more significant role as the shock intensity increases.

(2) Dislocation nucleation on the void surfaces is triggered by the precursor wave. The earliest appearing dislocations locate around a latitude circle of the spherical void surface. The latitudes of the earliest dislocation nucleation circle take values from $40°$ to $45°$. Dislocation nucleation mainly occurs in low latitude region of the void surface between the earliest dislocation nucleation circle and the equator. A continuum wave reflection theory and a resolved shear stress model are proposed to explain the distribution of the dislocation nucleation sites on the void surfaces. The simulation results are in reasonable agreement with the predictions of the continuum models.

(3) In nanoporous Al, after the passage of shock front, the temperature rises while the pressure drops. This is an abnormal phenomenon which is incompatible with the conventional Rankine-Hugonoit theory. In experiments, similar abnormalities for macroporous materials, for example pressure arising with density dropping, have been reported in the literature[16]. However, in MD simulations, such kind of thermodynamic abnormalities is captured in the present work for the first time. The abnormal phenomenon is explained by the nonequilibrium processes involved in void collapse.

(4) The differences in plastic relaxation between single-crystal Al and nanoporous Al mainly lie in the evolution of shear strain during release stage. In single-crystal Al, the shear strain reaches the saturated value after the passage of the compression front. The release wave result in decrease of shear strain due to partial recovery of the crystal slipping bands. The recovery of the shear slipping in single-crystal Al is due to lack of stable obstacles such as dislocation forests and junctions. In nanoporous Al, the shear strain accumulation under compression is attributed to the motion of dislocations nucleated from void surfaced. The dislocations are severely entangled and much obstacles are produced. As a result, the shear stress can not be relaxed by recovery of existing slipping bands, and has to be relaxed by producing new slipping bands. This results in increase of shear strain during release stage.

(5) The dynamical evolution processes of the dislocation structures in single-crystal Al and nonoporous Al under shock compression and release are studied. Shockley partial dislocations are nucleated after the sweeping of the shock wave front. The dislocation density reaches maximum after the passage of the shock wave front. During compression stage, the intersection of the Shockley partial dislocations on different slipping planes leads to formation of stair-rod dislocations. Then, release of compression loading leads to constriction of the partial dislocations to form perfect dislocations. Dislocation annihilation is apparently observed under compression and release in both single-crystal Al and nanoporous Al. During release stage, dislocation annihilation is faster in single-crystal Al than in nanoporous Al.

(6) The global spall fracture of the nanoporous Al is studied. Under same loading intensity, the spall strength of nanoporous Al is lower than that of single-crystal Al. This is because the void collapse in nanoporous Al leads to higher spall temperature; and higher temperature results in lower spall strength. On the other hand, the spall resistance, which is defined as the critical loading velocity that induces spall of the sample, is found to be higher for nanoporous Al than that for single-crystal Al. This is because that nanovoid collapse lead to attenuation of the stress and the attenuation is more obvious for higher porosity. The competition between the spall strength softening effect and the stress attenuation effect induced by the void collapse determines the difference in spall resistance between nanoporous Al and single-crystal Al. In our simulations, the stress attenuation effect prevails. Thus, the spall resistance of nanoporous Al is higher than that of single-crystal Al.

**Acknowledgements**

This work is supported by National Natural Science Foundation of China (No.11302052, No.11572053) and the Science Challenging Program of China.